\documentclass[runningheads]{llncs}
\usepackage[T1]{fontenc}
\usepackage{graphicx,verbatim}
\usepackage{multirow}
\usepackage{amssymb}
\usepackage{amsmath}
\usepackage{booktabs} 
\begin{document}
\title{VesselBridge3D: A Foundation Model Adaptation Framework for Label-Efficient 3D Vessel Segmentation}
\titlerunning{VesselBridge3D}
%

\author{Kirato Yoshihara\inst{1,2} \and
Yohei Sugawara\inst{2} \and
Yuta Tokuoka\inst{2} \and
Lihang Hong\inst{2}}

\authorrunning{K. Yoshihara et al.}

\institute{The University of Osaka \and
Preferred Networks, Inc. \\
\email{kiratoyoshihara@gmail.com, \{suga, tokuoka, rihankoo\}@preferred.jp}}
    
\maketitle              

\begin{abstract}
State-of-the-art vessel segmentation methods typically require large-scale annotated datasets and suffer from severe performance degradation under domain shifts. In clinical practice, however, acquiring extensive annotations for every new scanner or protocol is unfeasible. To address this, we propose VesselBridge3D, a foundation model adaptation framework that bridges frozen vision foundation models and volumetric vessel segmentation through lightweight 3D adaptation modules. The framework combines a lightweight 3D Adapter, a multi-scale 3D Aggregator, and Z-channel embedding for efficient adaptation to volumetric medical images. We instantiate VesselBridge3D with three frozen foundation encoders (DINOv3, MedSAM, and MedGemma) and evaluate it on the TopCoW (ID) and Lausanne (OOD) datasets. In the extreme low-data regime with 5 training samples, our method achieved a Dice score of 43.42\%, marking a 30$\%$ relative improvement over the state-of-the-art nnU-Net (33.41\%) and outperforming other Transformer-based baselines by up to 45\%. The proposed framework was effective across all evaluated frozen foundation encoders, with DINOv3 yielding the best performance in the most label-efficient settings. Furthermore, in the out-of-distribution setting, our model demonstrated superior robustness, achieving a 50$\%$ relative improvement over nnU-Net (21.37\% vs. 14.22\%), which suffered from severe domain overfitting. Ablation studies confirmed the effectiveness of the proposed 3D adaptation modules. Our results demonstrate that VesselBridge3D is an effective framework for label-efficient 3D vessel segmentation under data scarcity and domain shifts.


\end{abstract}

\section{Introduction}

Accurate segmentation of cerebrovascular structures is critical for diagnosing neurovascular diseases, planning stent placement, and managing intracranial aneurysms~\cite{MOCCIA201871,tetteh2020deepvesselnet,phellan2017vascular}. Despite the success of standard supervised deep learning in medical image analysis, state-of-the-art approaches, such as nnU-Net~\cite{isensee2021nnu}, rely heavily on large-scale, high-quality annotated datasets. In clinical practice, however, acquiring extensive voxel-level annotations for every new scanner, protocol, or modality is prohibitively labor-intensive and requires expert radiological knowledge~\cite{tajbakhsh2020embracing}.

Existing solutions for data-efficient learning, such as episodic Few-Shot Learning~\cite{ouyang2020self} and Unsupervised Domain Adaptation~\cite{kamnitsas2017unsupervised,dou2019pnp}, have attempted to mitigate this burden. Here, we focus on supervised label-efficient segmentation using a small number of fully annotated 3D volumes. However, standard supervised CNNs such as nnU-Net may overfit when trained with limited annotations, resulting in reduced generalization capability. Furthermore, they suffer from catastrophic performance degradation under out-of-distribution (OOD) shifts, where differences in imaging physics (e.g., MRI field strength, voxel spacing) cause the model to fail on unseen data~\cite{9557808,guan2021domain}. While Transformer-based architectures like SwinUNETR~\cite{hatamizadeh2022swin} offer better global context, they typically require even larger datasets to converge, exacerbating the data scarcity issue.

To address these challenges, we propose VesselBridge3D, a foundation model adaptation framework for robust 3D vessel segmentation. Our framework adapts frozen vision foundation models (e.g., DINOv3~\cite{simeoni2025dinov3}) through lightweight 3D adaptation modules.

While recent foundation model adaptations such as MedSAM~\cite{ma2024medsam}, SAM-Med3D~\cite{wang2023sammed3d}, and MedGemma~\cite{medgemma} have demonstrated the potential of foundation models for medical image analysis, we evaluate our generic foundation model adaptation framework on the challenging task of label-efficient vessel segmentation, where preserving inter-slice continuity is particularly important~\cite{phellan2017vascular,cldice2021}. We evaluate the proposed framework on the TopCoW dataset as the in-domain (ID) benchmark and the Lausanne dataset for out-of-distribution (OOD) evaluation. 

Our main contributions are summarized as follows:
\begin{enumerate}
    \item We propose VesselBridge3D, a foundation model adaptation framework for label-efficient 3D vessel segmentation using frozen vision foundation models.
    
    \item We introduce a lightweight 3D adaptation mechanism consisting of Z-channel embedding, a multi-scale 3D Aggregator, and a lightweight 3D Adapter to bridge 2D pre-training and 3D medical modalities.

    \item We evaluate VesselBridge3D on the TopCoW (ID) and Lausanne (OOD) datasets. The proposed framework outperforms nnU-Net by 30\% with only 5 annotated volumes and demonstrates a 50\% relative improvement in the OOD setting.
    
\end{enumerate}

\section{Method}
\label{sec:method}

\begin{figure}[t]
    \centering
    \includegraphics[width=\textwidth]{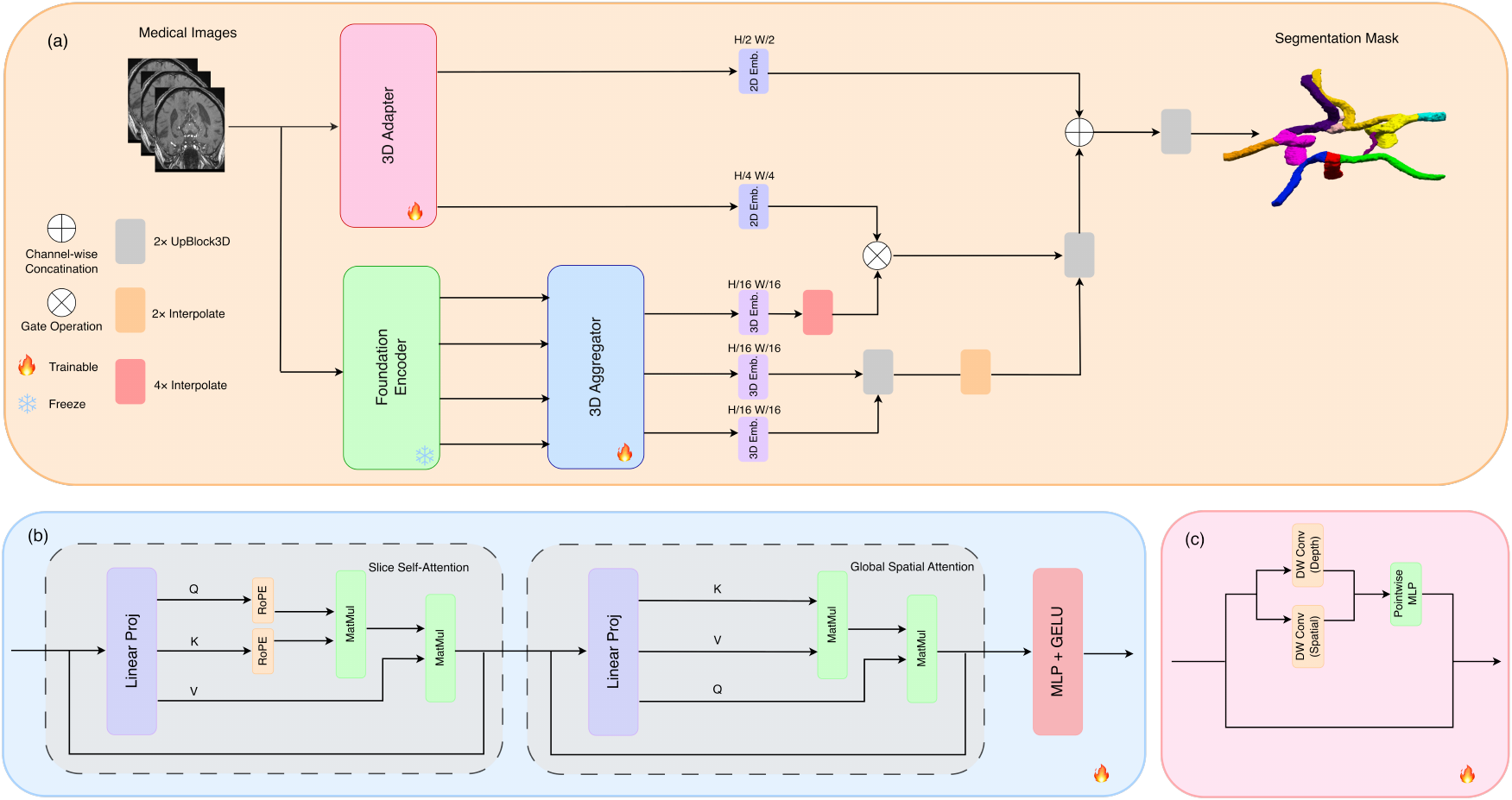}
    \caption{Overview of the proposed VesselBridge3D framework. (a) The overall architecture adapts a frozen vision foundation encoder to extract semantic features, while a parallel Lightweight 3D Adapter captures high-frequency volumetric details. (b) The 3D Aggregator fuses multi-scale features via factorized slice and global spatial attention. (c) The decoupled convolution block within the 3D Adapter efficiently models spatial and inter-slice dependencies.}
    \label{fig:overview}
\end{figure}

\label{sec:method-overview}
\subsection{Overview of the Framework}
As illustrated in Fig.~\ref{fig:overview}(a), VesselBridge3D adapts a frozen 2D vision foundation encoder for 3D segmentation via a side-tuning design~\cite{zhang2020side}. Unless otherwise specified, we instantiate the framework using DINOv3~\cite{simeoni2025dinov3}.

Given an input volume $X \in \mathbb{R}^{B \times 1 \times D \times H \times W}$, a frozen foundation backbone $\Phi_{frozen}$ and a 3D Aggregator $\mathcal{A}_{agg}$ extract global semantic context $F_{sem} = \mathcal{A}_{agg}(\Phi_{frozen}(\mathcal{T}(X)))$, where $\mathcal{T}(\cdot)$ applies slice-wise pseudo-coloring. In parallel, a trainable 3D Adapter $\Psi_{train}$ captures high-frequency volumetric details $F_{spat} = \Psi_{train}(X)$ directly from the raw input. The final prediction $\hat{Y} \in \mathbb{R}^{B \times 1 \times D \times H \times W}$ is generated by fusing these features via a gated mechanism and passing them to the decoder $\mathcal{D}_{dec}$:
\begin{equation}
    \hat{Y} = \mathcal{D}_{dec}(\text{Gate}(F_{sem}, F_{spat}))
\end{equation}
Only the lightweight modules ($\Psi_{train}$, $\mathcal{A}_{agg}$, and $\mathcal{D}_{dec}$) are trained, making the framework compatible with different frozen vision foundation encoders.



\subsection{Z-channel Embedding for 3D Awareness}
Frozen 2D encoders lack volumetric awareness. To inject slice location parameter-free, we form a slice-wise pseudo-color input with intensity in the R/G channels and a depth prior in B:
\begin{equation}
    X_{enc} = \text{Norm}(\text{Concat}(I_{gray}, I_{gray}, Z_{map})), \quad Z_{map} = z/D,
\end{equation}
where $I_{gray}$ is the per-slice min--max normalized intensity and $Z_{map} \in [0,1)$ is the relative slice depth broadcast over $H \times W$. Retaining intensity in the luminance channels preserves texture, while the B channel encodes depth as a per-slice bias.

\subsection{Shared Axial Aggregator and Multi-scale Fusion}
Unlike bottleneck-only foundation-model injection, the Shared Axial Aggregator (Fig.~\ref{fig:overview} (b)) fuses frozen features across scales to retain thin-vessel detail, following the alternating-attention design of VGGT~\cite{wang2025vggt}. Slice Self-Attention with RoPE~\cite{su2024roformer} is followed by Global Spatial Attention:
\[
H=\mathrm{MSA}_{Global}(\mathrm{MSA}_{Slice}(F)).
\]
The aggregated features are hierarchically upsampled and fused with the adapter features $F_{spat}$ through a gated mechanism:
\begin{equation}
    F_{out}=\text{Up}(H)+\sigma(W_{gate}\cdot\text{Concat}(\text{Up}(H),F_{spat}))\odot F_{spat}
\end{equation}
where $\sigma$ is the sigmoid function and $W_{gate}$ is a learnable convolution.

\subsection{Lightweight 3D Adapter}
To recover volumetric context, we introduce a parallel 3D CNN branch constructed from Anisotropic ConvNeXt Blocks~\cite{liu2022convnet}, as detailed in Fig.~\ref{fig:overview}(c). Instead of computationally expensive 3D convolutions, we decompose the depthwise convolution (DWConv) into parallel spatial and depth-wise branches. Formally, for an input $x_{in}$, the depthwise output $x_{dw}$ is:
\begin{equation}
    x_{dw} = \text{DWConv}_{3 \times 7 \times 7}(x_{in}) + \text{DWConv}_{3 \times 1 \times 1}(x_{in})
\end{equation}
The two branches capture spatial and inter-slice context, followed by pointwise convolutions with a residual connection:
\begin{equation}
    x_{out} = \text{PW}_{2}(\text{GELU}(\text{PW}_{1}(\text{LN}(x_{dw})))) + x_{in}
\end{equation}
where PW denotes $1\times1\times1$ convolution. This decoupled design efficiently extracts multi-scale features ($F_{spat}$) at $1/2$ and $1/4$ resolutions for fusion.

\section{Experiments and Results}
\label{sec:experiments}

\subsection{Datasets and Implementation Details}
\noindent\textbf{Datasets.} We validate our framework on two public cerebrovascular datasets, utilized in accordance with their open-access licenses without requiring additional IRB approval.

\noindent\textbf{TopCoW (ID):} We use the TopCoW challenge dataset~\cite{yang2024benchmarking}, which contains 125 MRA volumes with voxel-level annotations. The dataset is split into training ($N=87$), validation ($N=25$), and internal test ($N=13$) sets. To evaluate label efficiency under extreme data scarcity, we train using only 5 randomly sampled annotated volumes from the training set.

\noindent\textbf{Lausanne (OOD):} For OOD evaluation, we employ the Lausanne dataset (OpenNeuro ds003949)~\cite{dinoto2023towards} ($N=128$ TOF-MRA, aneurysm patients). Lacking manual segmentation, we binarized the probabilistic atlas~\cite{mouches2019} at an empirically optimized 0.2 threshold to preserve Circle of Willis connectivity. For fair evaluation, we restricted the ground truth to a central 60\% ROI, strictly matching TopCoW's physical Field-of-View.

\noindent\textbf{Implementation.} We instantiated VesselBridge3D using a frozen ViT-S/16 DINOv3 encoder (pre-trained on LVD-1689M) unless otherwise specified. For foundation model comparisons, we additionally evaluated the framework with frozen MedSAM and MedGemma encoders while keeping the proposed 3D adaptation modules, decoder, training protocol, and hyperparameters unchanged. For DINOv3, features were extracted from Layers \{2, 5, 8, 11\}; for MedSAM and MedGemma, corresponding intermediate layers were selected according to their architectures. All models were optimized using a compound loss of Soft Dice and Cross-Entropy. Input volumes were resized to $336 \times 336$ and randomly cropped to 64 depth slices, with standard augmentations (rotation, scaling, intensity shift) applied. For all foundation encoder comparisons, only the frozen encoder was replaced, while all other components and training settings remained identical.

\noindent\textbf{Baseline Configurations.} We compare VesselBridge3D against 3D U-Net~\cite{cicek20163d}, SwinUNETR~\cite{hatamizadeh2022swin}, UNETR~\cite{hatamizadeh2022unetr}, and nnU-Net (v2)~\cite{isensee2021nnu}. To ensure a fair comparison, all methods were trained using identical input resolutions ($336 \times 336$) and data augmentation strategies.

\noindent\textbf{Model Efficiency.} 
Unless otherwise specified, the reported model efficiency corresponds to the DINOv3 instantiation of VesselBridge3D, which employs a frozen ViT-S/16 distilled encoder (21.6M parameters).
Notably, thanks to our efficient side-tuning design, the number of trainable parameters is limited to only 13.6M. 
This is significantly more parameter-efficient compared to fully fine-tuning state-of-the-art 3D Transformers such as UNETR (122.3M) and SwinUNETR (62.2M), or even the standard nnU-Net (30.8M). 
The frozen encoder and compact adapter limit task-specific capacity, reducing overfitting risk versus end-to-end nnU-Net.

\subsection{Quantitative Analysis}
\label{sec:quantitative}

\begin{figure}[t]
    \centering
    \includegraphics[width=1.0\linewidth]{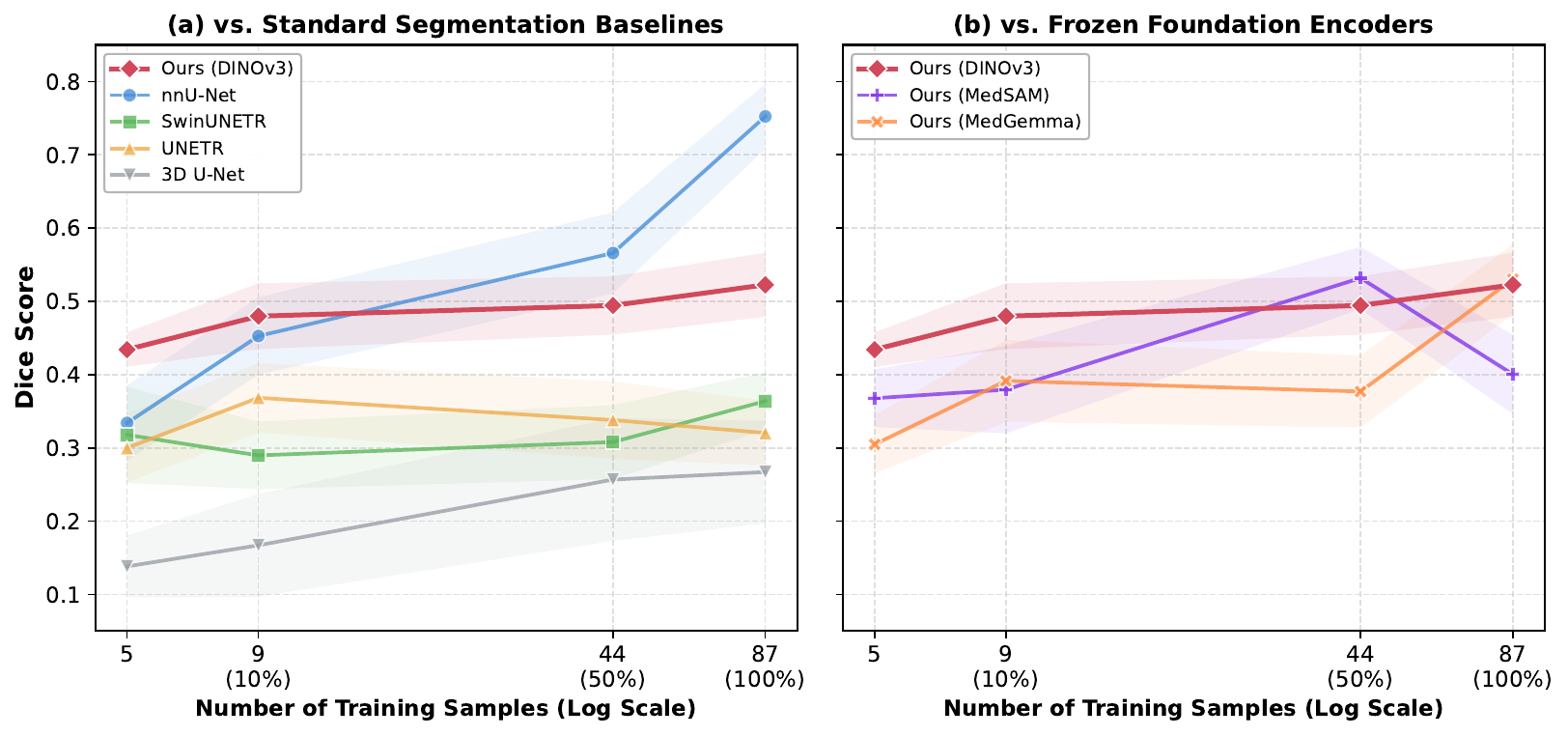}
    \caption{Label efficiency analysis on the TopCoW (ID) dataset. (a) Comparison with conventional segmentation baselines. Our method consistently outperforms nnU-Net, SwinUNETR, UNETR, and 3D U-Net in the low-data regime, achieving a 30.0\% relative improvement over nnU-Net with only five annotated volumes. (b) Comparison of VesselBridge3D instantiated with different frozen foundation encoders (DINOv3, MedSAM, and MedGemma). The proposed framework remains effective across all evaluated encoders, while DINOv3 provides the strongest performance in the most label-efficient settings. Shaded regions indicate the mean $\pm$ standard deviation of the Dice score across the 13 TopCoW test volumes for each trained model.}
    \label{fig:learning_curve}
\end{figure}

\noindent\textbf{In-domain Performance (TopCoW).}
We evaluated segmentation performance across training regimes ranging from extreme label scarcity ($N=5$) to full supervision ($N=87$). As illustrated in Fig.~\ref{fig:learning_curve}, our framework demonstrates remarkable label efficiency.
When trained using only 5 annotated volumes, our method achieves a Dice score of 43.42\%, significantly outperforming the strong baseline nnU-Net (33.41\%). This corresponds to a 30.0\% relative improvement, demonstrating the effectiveness of the proposed VesselBridge3D framework under severe label scarcity. Other Transformer baselines (SwinUNETR: 31.78\%, UNETR: 30.00\%) also perform less effectively, underscoring the difficulty of learning 3D representations from scratch using only a small number of annotated volumes. Among the evaluated frozen foundation encoders, DINOv3 achieved the strongest performance in the most label-efficient settings (5 volumes and 10\%), while MedSAM and MedGemma remained competitive under larger training regimes.

\noindent\textbf{Capacity-Efficiency Trade-off.}
We observe a performance cross-over as data availability increases. In the high-data regime ($N \ge 44$), nnU-Net eventually surpasses our method (75.24\% vs. 52.26\% at $N=87$). This behavior illustrates a fundamental trade-off: the frozen 2D encoder regularizes scarce-data learning but limits task-specific feature adaptation as labels increase, whereas fully trainable nnU-Net can adapt all features to exploit additional annotations.

\noindent\textbf{Quantitative Analysis on the OOD dataset.}
To assess generalization, we evaluated models trained on TopCoW (ID) directly on the unseen Lausanne dataset (OOD). Table~\ref{tab:ood_results} confirms that our method consistently outperforms nnU-Net across all data regimes.
When trained using only 5 annotated TopCoW volumes, our approach achieved a 50.3\% relative improvement in Dice and a 58.4\% gain in topological connectivity (clDice). This indicates that adapting frozen foundation encoders through VesselBridge3D provides robust transferable representations, particularly in low-data settings. Crucially, while increasing source training data caused nnU-Net to overfit (stagnant Dice, worsening HD95), our method continued to improve, widening the performance gap to a 63.8\% relative advantage in the fully supervised setting. These results demonstrate the robustness of VesselBridge3D under domain shifts.

\begin{table}[t]
\centering
\caption{Cross-domain generalization results on the Lausanne dataset (OOD). VesselBridge3D (DINOv3 instantiation) outperforms nnU-Net in all metrics across all data regimes. While nnU-Net degrades in shape fidelity (HD95 increases) as training data increases, our method maintains robust performance, highlighting the generalizability of the frozen backbone.}
\label{tab:ood_results}
\renewcommand{\arraystretch}{1.1} 
\setlength{\tabcolsep}{5pt}

\fontsize{8}{9.5}\selectfont 

\begin{tabular}{lcccccc}
\toprule
\multirow{2}{*}{\textbf{Data}} & \multicolumn{3}{c}{\textbf{nnU-Net}} & \multicolumn{3}{c}{\textbf{Ours (DINOv3)}} \\
\cmidrule(lr){2-4} \cmidrule(lr){5-7}
 & Dice(\%) $\uparrow$ & clDice $\uparrow$ & HD95(mm) $\downarrow$ & \textbf{Dice(\%)} $\uparrow$ & \textbf{clDice} $\uparrow$ & \textbf{HD95(mm)} $\downarrow$ \\
\midrule
5 volumes & 14.22 & 0.053 & 22.61 & \textbf{21.37} & \textbf{0.084} & \textbf{15.94} \\
10\%   & 14.89 & 0.051 & 24.37 & \textbf{17.19} & \textbf{0.076} & \textbf{16.45} \\
50\%   & 11.23 & 0.037 & 31.25 & \textbf{24.05} & \textbf{0.108} & \textbf{14.21} \\
100\%  & 13.74 & 0.057 & 29.03 & \textbf{22.50} & \textbf{0.095} & \textbf{15.77} \\
\bottomrule
\end{tabular}
\end{table}

\subsection{Qualitative Analysis}
\label{sec:qualitative}

\begin{figure}[t]
    \centering
    \includegraphics[width=\textwidth]{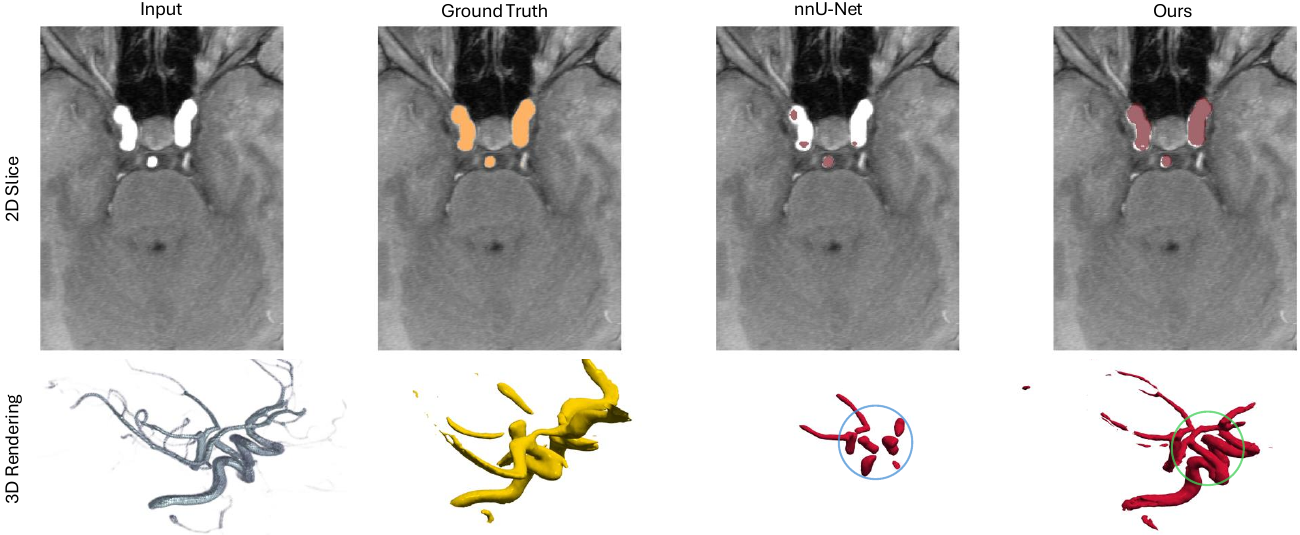}
    \caption{Qualitative comparison on the unseen OOD Lausanne dataset using models trained on 5 annotated TopCoW volumes. Samples were randomly selected to ensure unbiased evaluation. Top Row: A 2D slice where nnU-Net fails due to domain shift, while VesselBridge3D (DINOv3) correctly delineates the vessel. Bottom Row: 3D rendering of a different subject; the baseline produces fragmented results (blue circle), while VesselBridge3D (DINOv3) preserves vascular connectivity (green circle).}
    \label{fig:qualitative}
\end{figure}

Visual comparisons on the external Lausanne dataset (OOD) are presented in Fig.~\ref{fig:qualitative}. To assess cross-domain robustness, we randomly selected a Lausanne case and evaluated models trained using only 5 annotated volumes from the TopCoW dataset.

\noindent\textbf{Robustness to Domain Shift (2D View).}
Fig.~\ref{fig:qualitative}(Top Row) displays representative 2D slices. As shown, the baseline nnU-Net struggles to generalize to the target domain, failing to detect the vascular structure evident in the input. This indicates that standard supervised learning overfits to the ID intensity distribution when training data is scarce. In contrast, our method accurately identifies the vessel boundaries. This demonstrates that VesselBridge3D effectively transfers robust representations from frozen foundation encoders, maintaining performance under significant domain shifts.

\noindent\textbf{Topological Integrity (3D View).}
The limitation of the baseline is further amplified in the 3D reconstructions in Fig.~\ref{fig:qualitative}(Bottom Row). Due to the lack of robust 2D detection, nnU-Net generates severe discontinuities and isolated fragments (blue circle). Conversely, VesselBridge3D (DINOv3) successfully preserves the topological connectivity of the vessel tree (green circle). This demonstrates that the proposed 3D adaptation framework effectively maintains global structural coherence while improving pixel-level accuracy.

\subsection{Ablation Study}

\begin{table}[t]
\centering
\caption{Ablation study of VesselBridge3D (DINOv3) on the TopCoW dataset using 5 annotated training volumes. We evaluate the contribution of key components: 3D Adapter, 3D Aggregator, Multi-scale fusion, and Z-channel. The full model yields the best balance of volumetric overlap (Dice\%) and topological preservation (clDice), while maintaining boundary accuracy (HD95) within 1.5 mm of the best variant.}
\label{tab:ablation}
\renewcommand{\arraystretch}{1.1}
\setlength{\tabcolsep}{4pt}

\fontsize{8}{9.5}\selectfont

\begin{tabular}{l|ccc}
\toprule
\textbf{Method} & \textbf{Dice(\%)} $\uparrow$ & \textbf{clDice} $\uparrow$ & \textbf{HD95(mm)} $\downarrow$ \\ 
\midrule
w/o 3D Aggregator & 22.35 & 0.279  & 60.97 \\
w/o 3D Adapter & 28.67  & 0.072 & 85.33  \\
Only Last Layer (11) & 42.88  & 0.284  & 53.37  \\
Only Last Layer (11) w/o Z & 41.10 & 0.304  & \textbf{52.87} \\
\midrule
\textbf{VesselBridge3D (Full)} & \textbf{43.42}  & \textbf{0.310}  & 54.29 \\
\bottomrule
\end{tabular}
\end{table}

Table~\ref{tab:ablation} validates the proposed VesselBridge3D framework. Removing the 3D Adapter or 3D Aggregator causes substantial performance degradation, demonstrating that lightweight 3D adaptation is essential for effectively transferring frozen foundation encoders to volumetric segmentation. Furthermore, multi-scale aggregation and Z-channel embedding consistently improve segmentation accuracy.

\section{Discussion and Conclusion}
In this study, we presented VesselBridge3D, a foundation model adaptation framework for label-efficient volumetric vessel segmentation. Our experiments demonstrated that lightweight 3D adaptation effectively transfers frozen vision foundation encoders to volumetric medical imaging, outperforming supervised baselines in low-data regimes and improving robustness under domain shift. While fully fine-tuned models eventually surpass the proposed framework with abundant annotations due to the capacity limits of frozen encoders, VesselBridge3D substantially reduces the dependence on large-scale annotated datasets. These results suggest that VesselBridge3D provides an effective framework for label-efficient volumetric medical image segmentation.

%
%
%
\bibliographystyle{splncs04}
\bibliography{reference}

@article{isensee2021nnu,
  title={nnU-Net: a self-configuring method for deep learning-based biomedical image segmentation},
  author={Isensee, Fabian and Jaeger, Paul F. and Kohl, Simon A. A. and Petersen, Jens and Maier-Hein, Klaus H.},
  journal={Nature Methods},
  volume={18},
  pages={203--211},
  year={2021},
  publisher={Nature Publishing Group},
  doi={10.1038/s41592-020-01008-z}
}

@article{MOCCIA201871,
  title = {Blood vessel segmentation algorithms — Review of methods, datasets and evaluation metrics},
  journal = {Computer Methods and Programs in Biomedicine},
  volume = {158},
  pages = {71--91},
  year = {2018},
  issn = {0169-2607},
  doi = {10.1016/j.cmpb.2018.02.001}, 
  author = {Sara Moccia and Elena {De Momi} and Sara {El Hadji} and Leonardo S. Mattos}
}

@ARTICLE{9557808,
  author={Guan, Hao and Liu, Mingxia},
  journal={IEEE Transactions on Biomedical Engineering}, 
  title={Domain Adaptation for Medical Image Analysis: A Survey}, 
  year={2022},
  volume={69},
  pages={1173-1185},
  doi={10.1109/TBME.2021.3117407}}

@inproceedings{hatamizadeh2022swin,
  title={Swin unetr: Swin transformers for semantic segmentation of brain tumors in mri images},
  author={Hatamizadeh, Ali and Nath, Vishwesh and Tang, Yucheng and Yang, Dong and Roth, Holger R and Xu, Daguang},
  booktitle={International MICCAI brainlesion workshop},
  pages={272--284},
  year={2021},
  organization={Springer}
}

@article{simeoni2025dinov3,
  title={DINOv3},
  author={Siméoni, Oriane and Vo, Huy V. and Seitzer, Maximilian and Baldassarre, Federico and Oquab, Maxime and Jose, Cijo and Khalidov, Vasil and Szafraniec, Marc and Yi, Seungeun and Ramamonjisoa, Michaël and others},
  journal={arXiv preprint arXiv:2508.10104},
  year={2025},
  url={https://arxiv.org/abs/2508.10104}
}

@Article{liu2022convnet,
  author  = {Zhuang Liu and Hanzi Mao and Chao-Yuan Wu and Christoph Feichtenhofer and Trevor Darrell and Saining Xie},
  title   = {A ConvNet for the 2020s},
  journal = {Proceedings of the IEEE/CVF Conference on Computer Vision and Pattern Recognition (CVPR)},
  year    = {2022},
}

@article{su2024roformer,
  title={RoFormer: Enhanced Transformer with Rotary Position Embedding},
  author={Su, Jianlin and Lu, Yu and Pan, Shengfeng and Murtadha, Ahmed and Wen, Bo and Liu, Yunfeng},
  journal={Neurocomputing},
  volume={568},
  pages={127063},
  year={2024},
  publisher={Elsevier}
}

@article{dinoto2023towards,
  title={Towards Automated Brain Aneurysm Detection in {TOF-MRA}: Open Data, Weak Labels, and Anatomical Knowledge},
  author={Di Noto, T. and Marie, G. and Tourbier, S. and Alem{\'a}n-G{\'o}mez, Y. and Esteban, O. and Saliou, G. and Cuadra, M. B. and Hagmann, P. and Richiardi, J.},
  journal={Neuroinformatics},
  volume={21},
  pages={21--34},
  year={2023},
  publisher={Springer},
  doi={10.1007/s12021-022-09597-0}
}

@article{yang2024benchmarking,
  title={Benchmarking the {CoW} with the {TopCoW} Challenge: Topology-Aware Anatomical Segmentation of the Circle of Willis for {CTA} and {MRA}},
  author={Yang, Kaiyuan and Musio, Fabio and Ma, Yihui and others},
  journal={arXiv preprint arXiv:2312.17670},
  year={2023}
}

@article{mouches2019,
  title={A statistical atlas of cerebral arteries generated using multi-center {MRA} datasets from healthy subjects},
  author={Mouches, Pauline and Forkert, Nils D.},
  journal={Scientific Data},
  volume={6},
  pages={29},
  year={2019},
  publisher={Nature Publishing Group},
  doi={10.1038/s41597-019-0034-5}
}

@inproceedings{hatamizadeh2022unetr,
  title={Unetr: Transformers for 3d medical image segmentation},
  author={Hatamizadeh, Ali and Tang, Yucheng and Nath, Vishwesh and Yang, Dong and Myronenko, Andriy and Landman, Bennett and Roth, Holger R and Xu, Daguang},
  booktitle={Proceedings of the IEEE/CVF winter conference on applications of computer vision},
  pages={574--584},
  year={2022}
}

@inproceedings{cicek20163d,
  title={3D U-Net: learning dense volumetric segmentation from sparse annotation},
  author={{\c{C}}i{\c{c}}ek, {\"O}zg{\"u}n and Abdulkadir, Ahmed and Lienkamp, Soeren S and Brox, Thomas and Ronneberger, Olaf},
  booktitle={International conference on medical image computing and computer-assisted intervention},
  pages={424--432},
  year={2016},
  organization={Springer}
}

@inproceedings{wang2025vggt,
  title={VGGT: Visual Geometry Grounded Transformer},
  author={Wang, Jianyuan and Chen, Minghao and Karaev, Nikita and Vedaldi, Andrea and Rupprecht, Christian and Novotny, David},
  booktitle={Proceedings of the IEEE/CVF Conference on Computer Vision and Pattern Recognition},
  year={2025}
}

@article{tetteh2020deepvesselnet,
  author  = {Tetteh, Giles and Efremov, Velizar and Forkert, Nils D. and Schneider, Matthias and Kirschke, Jan and Weber, Bruno and Zimmer, Claus and Piraud, Marie and Menze, Bj{\"o}rn H.},
  title   = {DeepVesselNet: Vessel Segmentation, Centerline Prediction, and Bifurcation Detection in 3-D Angiographic Volumes},
  journal = {Frontiers in Neuroscience},
  volume  = {14},
  year    = {2020},
  doi     = {10.3389/fnins.2020.592352},
  issn    = {1662-453X}
}

@article{tajbakhsh2020embracing,
  title={Embracing imperfect datasets: A review of deep learning solutions for medical image segmentation},
  author={Tajbakhsh, Nima and Jeyaseelan, Laura and Li, Qian and Chiang, Jeffrey N and Wu, Zhihao and Ding, Xiaowei},
  journal={Medical Image Analysis},
  volume={63},
  pages={101693},
  year={2020},
  publisher={Elsevier},
  url={https://doi.org/10.1016/j.media.2020.101693}
}

@inproceedings{ouyang2020self,
  title={Self-supervision with superpixels: Training few-shot medical image segmentation without annotation},
  author={Ouyang, Cheng and Biffi, Carlo and Chen, Chen and Kart, Turkay and Qiu, Huaqi and Rueckert, Daniel},
  booktitle={European conference on computer vision},
  pages={762--780},
  year={2020},
  organization={Springer}
}

@inproceedings{kamnitsas2017unsupervised,
  author    = {Kamnitsas, Konstantinos and Baumgartner, Christian and Ledig, Christian and Newcombe, Virginia and Simpson, Joanna and Kane, Andrew and Menon, David and Nori, Aditya and Criminisi, Antonio and Rueckert, Daniel and Glocker, Ben},
  editor    = {Niethammer, Marc and Styner, Martin and Aylward, Stephen and Zhu, Hongtu and Oguz, Ipek and Yap, Pew-Thian and Shen, Dinggang},
  title     = {Unsupervised Domain Adaptation in Brain Lesion Segmentation with Adversarial Networks},
  booktitle = {Information Processing in Medical Imaging},
  year      = {2017},
  publisher = {Springer International Publishing},
  address   = {Cham},
  pages     = {597--609},
  isbn      = {978-3-319-59050-9}
}

@article{dou2019pnp,
  title={Pnp-adanet: Plug-and-play adversarial domain adaptation network at unpaired cross-modality cardiac segmentation},
  author={Dou, Qi and Ouyang, Cheng and Chen, Cheng and Chen, Hao and Glocker, Ben and Zhuang, Xiahai and Heng, Pheng-Ann},
  journal={Ieee Access},
  volume={7},
  pages={99065--99076},
  year={2019},
  publisher={IEEE}
}

@article{guan2021domain,
   title={Domain Generalization for Medical Image Analysis: A Review},
   author={Yoon, Jee Seok and Oh, Kwanseok and Shin, Yooseung and Mazurowski, Maciej A. and Suk, Heung-Il},
   journal={Proceedings of the IEEE},
   volume={112},
   number={10},
   pages={1583--1609},
   year={2024},
   publisher={IEEE},
   doi={10.1109/JPROC.2024.3507831}
}

@article{ma2024medsam,
  title={Segment anything in medical images},
  author={Ma, Jun and He, Yuting and Li, Feifei and Han, Lin and You, Cen and Wang, Bo},
  journal={Nature Communications},
  volume={15},
  pages={654},
  year={2024},
  publisher={Nature Publishing Group},
  url={https://www.nature.com/articles/s41467-024-44824-z}
}

@misc{wang2023sammed3d,
      title={SAM-Med3D: Towards General-purpose Segmentation Models for Volumetric Medical Images}, 
      author={Haoyu Wang and Sizheng Guo and Jin Ye and Zhongying Deng and Junlong Cheng and Tianbin Li and Jianpin Chen and Yanzhou Su and Ziyan Huang and Yiqing Shen and Bin Fu and Shaoting Zhang and Junjun He and Yu Qiao},
      year={2024},
      eprint={2310.15161},
      archivePrefix={arXiv},
      primaryClass={cs.CV},
      url={https://arxiv.org/abs/2310.15161}, 
}

@inproceedings{phellan2017vascular,
  title={Vascular Segmentation in {TOF MRA} Images of the Brain Using a Deep Convolutional Neural Network},
  author={Phellan, Renzo and Peixinho, Alan and Falc{\~a}o, Alexandre and Forkert, Nils D.},
  booktitle={Intravascular Imaging and Computer Assisted Stenting, and Large-Scale Annotation of Biomedical Data and Expert Label Synthesis},
  pages={39--46},
  year={2017},
  publisher={Springer International Publishing},
  address={Cham},
  doi={10.1007/978-3-319-67534-3_5}
}

@misc{zhang2020side,
      title={Side-Tuning: A Baseline for Network Adaptation via Additive Side Networks}, 
      author={Jeffrey O. Zhang and Alexander Sax and Amir Zamir and Leonidas Guibas and Jitendra Malik},
      year={2020},
      eprint={1912.13503},
      archivePrefix={arXiv},
      primaryClass={cs.LG},
      url={https://arxiv.org/abs/1912.13503}, 
}

@inproceedings{cldice2021,
  title={clDice-a Novel Topology-Preserving Loss Function for Tubular Structure Segmentation},
  author={Shit, Suprosanna and Paetzold, Johannes C and Sekuboyina, Anjany and Ezhov, Ivan and Unger, Alexander and Zhylka, Andrey and Pluim, Josien PW and Bauer, Ulrich and Menze, Bjoern H},
  booktitle={Proceedings of the IEEE/CVF Conference on Computer Vision and Pattern Recognition},
  pages={16560--16569},
  year={2021}
}

@article{medgemma,
  title={Medgemma technical report},
  author={Sellergren, Andrew and Kazemzadeh, Sahar and Jaroensri, Tiam and Kiraly, Atilla and Traverse, Madeleine and Kohlberger, Timo and Xu, Shawn and Jamil, Fayaz and Hughes, C{\'\i}an and Lau, Charles and others},
  journal={arXiv preprint arXiv:2507.05201},
  year={2025}
}
%






\end{document}